\documentclass[%
prd,
nofootinbib,
superscriptaddress,
]{revtex4}
\usepackage{amsmath}
\usepackage{amssymb}
\usepackage{braket}
\usepackage{bm}
\usepackage{color}
\usepackage{graphicx}
\begin{document}
\title{Observable cosmological vector mode in the dark ages}
\author{Shohei Saga}
\affiliation{Department of Physics and Astrophysics, Nagoya University,
Aichi 464-8602, Japan}
\email{saga.shohei@nagoya-u.jp}
\begin{abstract}
The second-order vector mode is inevitably induced from the coupling of first-order scalar modes in cosmological perturbation theory
and might hinder a possible detection of primordial gravitational waves from inflation through 21cm lensing observations.
Here, we investigate the weak lensing signal in 21cm photons emitted by neutral hydrogen atoms in the dark ages induced by the second-order vector mode by decomposing the deflection angle of the 21cm lensing signal into the gradient and curl modes.
The curl mode is a good tracer of the cosmological vector and tensor modes since the scalar mode does not induce the curl one.
By comparing angular power spectra of the 21cm lensing curl mode induced by the second-order vector mode and primordial gravitational waves whose amplitude is parametrized by the tensor-to-scalar ratio $r$, we find that the 21cm curl mode from the second-order vector mode dominates over that from primordial gravitational waves on almost all scales if $r \lesssim 10^{-5}$.
If we use the multipoles of the power spectrum up to $\ell_{\rm max} = 10^{5}$ and $10^{6}$ in reconstructing the curl mode from 21cm temperature maps, the signal-to-noise ratios of the 21cm curl mode from the second-order vector mode achieve ${\rm S/N} \approx 0.46$ and $73$, respectively.
Observation of 21cm radiation is, in principle, a powerful tool to explore not only the tensor mode but also the cosmological vector mode.
\end{abstract}
\maketitle
\section{introduction}
Recent development of cosmological observations plays an important role in establishing standard cosmology, namely, the $\Lambda$CDM model.
In the current states of observations of cosmological perturbations, for example,
we have detected the cosmic microwave background (CMB) temperature anisotropy, the CMB E-mode polarization, galaxy clustering,
and so on \cite{Tegmark:2006az,Hinshaw:2012aka,Sanchez:2013tga,Ade:2015xua}.
In the context of the cosmological perturbation theory, the perturbations can be decomposed into the scalar, vector, and tensor modes.
Current observations show good agreement with perturbations of the first-order scalar mode.
The vector and tensor modes are becoming the next observational targets.

The vector and tensor modes have been considered in many contexts.
The standard model of inflation in the early Universe, namely, the single-field slow-roll inflation, predicts the existence of primordial gravitational waves (PGWs) that correspond to the tensor mode.
The vector mode can be induced by cosmological magnetic fields \cite{Lewis:2004ef,Lewis:2004kg}, cosmological defects \cite{Pen:1997ae,Durrer:1998rw,Horiguchi:2015xsa}, or additional vector fields \cite{Zuntz:2010jp,Saga:2013glg}.
However, the amplitude of the vector mode generated in these models highly depends on their model parameters.
When we expand cosmological perturbations up to the second order,
the nonlinear coupling of the first-order scalar modes naturally induces the second-order vector and tensor modes \cite{Assadullahi:2009jc,Ananda:2006af,Saga:2014jca,Baumann:2007zm,Saga:2015apa,Ichiki:2006cd,Fenu:2010kh,Saga:2015bna}.
Since the amplitude of the first-order scalar mode is precisely determined by recent observations,
the amplitudes of second-order vector and tensor modes can be predicted without introducing additional model parameters.
The second-order mode is one of the good observational targets since those modes always exist.
Furthermore, the detection of cosmological vector or tensor modes is important to test the validity of the scalar, vector, and tensor decomposition.

Weak lensing is one of the tools that can be used to study the cosmological vector or tensor modes.
CMB photons are deflected by foreground perturbations such as density fluctuations, the gravitational wave background, or vector perturbations.
We can split the deflection pattern into the gradient (parity-even) and curl (parity-odd) modes \cite{Namikawa:2011cs,Yamauchi:2012bc,Yamauchi:2013fra}.
The curl mode is induced only from the vector and tensor modes.
Therefore, the curl mode is one of the good tracers to explore the cosmological vector or tensor modes.

In our previous study \cite{Saga:2015apa},
we discussed the detectability of the second-order curl mode in the CMB lensing and cosmic shear.
Unfortunately, because the signal from the second-order curl mode is small,
we concluded that we could not detect the second-order curl mode even with an ideal experiment for the full sky without the instrumental noise if we utilize the quadratic estimator method.
However, we found that the curl mode from the second-order vector mode is comparable to that from PGWs with tensor-to-scalar ratio $r< 0.1$, especially so in lower redshifts because the second-order vector mode is continuously generated, while PGWs always decay in time.
In other words, when there is an observation that enables us to detect PGWs with $r< 0.1$ through lensing,
we can also detect the second-order vector mode.

In previous studies \cite{Book:2011dz,Masui:2010cz,Sigurdson:2005cp}, it was shown that the 21cm lensing has a possibility, in principle, to detect PGWs with a quite small tensor-to-scalar ratio.
Long before reionization begins, no astronomical objects exist, and this era is called the dark ages.
Neutral hydrogen atoms emit 21cm line radiation that originates from the hyperfine structure; see, e.g., Ref.~\cite{Furlanetto:2006jb}.
In principle, we can observe the 21cm radiation from the redshift $z\approx 200$ to $30$ in future experiments.
21cm photons are deflected by the foreground scalar, vector, and tensor modes.
Moreover, we can decompose the deflection angle of the 21cm photons into the gradient and curl modes depending on the parity.
Compared with CMB fluctuations, 21cm radiation does not suffer from a diffusion mechanism such as Silk dumping
and the 21cm fluctuations on small scales remain until today.
Consequently, the available information from 21cm fluctuations is dramatically improved compared with that from CMB fluctuations.
Furthermore, 21cm radiation is emitted from each redshift and many maps are available.
For the above reason, 21cm lensing reconstruction noise would become quite small compared with CMB lensing reconstruction noise.
Therefore, although second-order vector and tensor signals tend to be small, there is a possibility to detect these second-order signals in 21cm lensing.

In this paper, we focus on the 21cm lensing curl mode induced from the second-order vector mode.
Our aim is to estimate the signal-to-noise ratio of the 21cm curl mode from the second-order vector mode in ideal experiments.
In standard cosmology, the first-order vector mode always decays and is neglected in linear theory.
The detection of the cosmological vector mode is quite important because it would become a proof of the cosmological perturbation theory itself and the validity of the scalar, vector, and tensor decomposition.

This paper is organized as follows.
This study is based on the physics of 21cm radiation, cosmological perturbation theory expanded up to second order, and weak lensing.
In Sec.~\ref{sec: pre}, we briefly review these physics and their mathematical basics with appropriate references.
To predict the 21cm radiation fluctuation, we need to solve the perturbed Boltzmann equation for the 21cm photons.
The Boltzmann equation for 21cm photons has the collision term with neutral hydrogen atoms.
We solve the Boltzmann equation for 21cm photons numerically.
The second-order vector mode is generated from the coupling of the first-order scalar potentials and is solved numerically in a straightforward manner.
To discuss the detectability, we derive the noise spectrum induced from the lensing reconstruction in a similar way as in CMB lensing reconstruction.
The advantage of 21cm lensing reconstruction is that one can coadd different redshift slices.
In Sec.~\ref{sec: results}, we show our main results and give some discussions regarding the 21cm lensing signal and detectability.
Finally, we present our conclusions in Sec.~\ref{sec: summary}.

\section{Preliminary}\label{sec: pre}
In this section, we quickly review 21cm physics \cite{Lewis:2007kz}, the second-order vector mode \cite{Saga:2015apa}, and the weak lensing curl mode \cite{Namikawa:2011cs}.
Throughout this paper, our calculation is based on a flat $\Lambda$CDM model with the standard six cosmological parameters constrained by Planck \cite{Ade:2015xua}.
We work in the Poisson gauge as
\begin{eqnarray}
ds^{2} = a^{2}(\eta) \left[ -\left( 1 + 2\Psi \right)d\eta^{2} + 2\sigma_{i}d\eta dx^{i} + \left( \delta_{ij} - 2\Phi \delta_{ij} + h_{ij}\right)dx^{i}dx^{j}\right] ~,
\end{eqnarray}
where $\eta$ is the conformal time.
Moreover, due to the gauge condition, the metric perturbations for the $(0,i)$ and $(i,j)$ components obey $\sigma^{i}{}_{,i} = \chi^{i}{}_{i} = \chi^{ij}{}_{,i} = 0$.
Note that we obey the rule that subscripts and superscripts of greek and latin characters starting from $i, j, \cdots$ run from 0 to 3 and from 1 to 3, respectively.

\subsection{21cm radiation from the dark ages}
Throughout this paper, we focus on the redshift $z \gtrsim 30$ since we are interested in the weak lensing signals from the dark ages.
During the dark ages, we can ignore the effect of Ly$\alpha$ photons emitted from astronomical objects.
Neutral hydrogen atoms form after recombination ($z\approx 1100$), and the effect of Ly$\alpha$ photons from stars dominates on the evolution of neutral hydrogen atoms after $z\approx 30$.
However, during $200\lesssim z \lesssim 1100$, thermal coupling between residual electrons and CMB photons brings the spin temperature of hydrogen atoms to the CMB temperature, and therefore no 21cm signal comes from this period.
Consequently, we can observe 21cm radiation during $30 \lesssim z \lesssim 200$.

The Boltzmann equation for 21cm photons $f_{21}(\eta,\bm{x},\epsilon, \bm{\hat{n}})$ can be written as
\begin{eqnarray}
\frac{df_{21}}{d\lambda} = C_{\rm H}[f_{21}] ~, \label{eq: Boltz 21cm}
\end{eqnarray}
where $\lambda$, $\epsilon$, and $\bm{\hat{n}}$ are the affine parameter, the energy of 21cm photons, and the direction of 21cm photons, respectively.
$C_{\rm H}[f_{21}]$ is the collision term due to the 21cm interaction.
The 21cm interaction includes the excitation due to the absorption of photons, deexcitation due to the spontaneous emission, and deexcitation due to the stimulated emission.
A detailed expression of the collision term is given in Ref.~\cite{Lewis:2007kz}.

We can solve the Boltzmann equation (\ref{eq: Boltz 21cm}) perturbatively, namely, $f_{21}(\eta,\bm{x},\epsilon, \bm{\hat{n}}) = f^{(0)}_{21}(\eta,\epsilon) + \delta f^{(1)}_{21}(\eta ,\bm{x},\epsilon, \bm{\hat{n}})$.
By solving the background Universe, we obtain the zeroth-order solution as
\begin{equation}
\delta T^{(0)}_{\rm b} = \left( 1 - e^{-\tau^{(0)}}\right)\left[ \frac{T^{(0)}_{\rm s} - T^{(0)}_{\rm CMB}}{1+z}\right] ~,
\end{equation}
where $\tau$, $T_{\rm CMB}$, and $T_{\rm s}$ are the optical depth of 21cm radiation, the CMB temperature, and the spin temperature, respectively.
The spin temperature is defined by the ratio of the number density in the upper states to the lower ones as
\begin{equation}
\frac{n_{1}}{n_{0}} = \frac{g_{1}}{g_{0}}\exp{\left( -\frac{\Delta E_{10}}{k_{\rm B}T_{\rm s}}\right)} ~,
\end{equation}
where $n_{1,0}$, $g_{1,0}$, and $\Delta E_{10}$ are the number density for each state, the number of degenerate states for each state, and the energy difference between the upper and lower states, respectively.
In the case of the hyperfine structure of neutral hydrogen atoms, $g_{1} = 3$, $g_{0} = 1$, and $\Delta E_{21} = h_{\rm p}\nu_{\rm 21cm}$, where $\nu_{\rm 21cm} \approx 1420$MHz.

The perturbed distribution function of 21cm photons is expanded by spherical harmonics as
\begin{equation}
a^{\rm 21cm}_{\ell,m}(z_{\epsilon}) = \int{d\Omega Y^{*}_{\ell , m}(\bm{\hat{n}}) \delta f^{(1)}_{21}(\eta_{\rm O},\bm{x}_{\rm O},\epsilon, \bm{\hat{n}})} ~,
\end{equation}
where $Y_{\ell,m}(\bm{\hat{n}})$ is the spin-0 spherical harmonics and $z_{\epsilon}$ is the redshift at which the 21cm photon was emitted.
The subscript ${\rm O}$ stresses that the values are evaluated at the observer.
By using the above coefficients, we can write down the angular power spectrum of the brightness temperature fluctuation as
\begin{eqnarray}
\Braket{a^{\rm 21 cm}_{\ell_{1},m_{1}}(z_{1}) a^{\rm 21 cm}_{\ell_{2},m_{2}}(z_{2})}
&\equiv& C^{\rm 21 cm}_{\ell_{1}}(z_{1},z_{2})(-1)^{m_{1}}\delta_{\ell_{1},\ell_{2}}\delta_{m_{1}, -m_{2}} ~.
\end{eqnarray}
To calculate the angular power spectrum of the brightness temperature fluctuation, we use the public code subroutine: the CAMB sources \cite{Lewis:2007kz}.
The feature of the 21cm power spectrum $C^{\rm 21cm}_{\ell}(z_{1},z_{2})$ is well discussed in \cite{Lewis:2007kz}, and it was found that there is no dumping in the 21cm power spectrum like the Silk dumping in the CMB power spectrum.
For this reason, the available maximum multipole in the 21cm power spectrum can reach $\ell_{\rm max} \approx 10^{6}\sim 10^{7}$, which corresponds to the Jeans scales.
Compared with the CMB temperature power spectrum, the available mode increases by $10^{3}\sim 10^{4}$.
Furthermore, by varying the observing frequency, we can take a number of redshift slices.
These advantages in using the 21cm power spectrum help us to decrease 21cm lensing reconstruction noise.

\subsection{Second-order vector mode}
In this subsection, we review the second-order vector mode induced from the first-order scalar modes following Ref.~\cite{Saga:2015apa}.
In standard cosmology at the linear perturbation level, the vector mode decays as $a^{-2}$.
For this reason, the vector mode is negligible in standard cosmology.
However, when we expand cosmological perturbation theory up to the second order,
the second-order vector mode appears from the nonlinearity.

The nonlinearity also naturally induces not only the second-order vector but also the tensor mode.
However, the dominant second-order mode on the weak lensing signal is the vector mode rather than the tensor mode \cite{Saga:2015apa}.
In the study of N-body simulations \cite{Bruni:2013mua,Thomas:2014aga,Adamek:2015mna}, even when we treat the higher-order tensor mode nonperturbatively by using N-body simulations, the amplification of the fully nonlinear effect is still negligible.
Therefore, in this paper, we focus not on the second-order tensor but on the vector mode.
For the purpose of comparison, we show our results with the weak lensing signal from PGWs parametrized by a tensor-to-scalar ratio $r$.

From here, we derive the evolution equation for the second-order vector mode.
The vector metric perturbation $\sigma_{i}$ is decomposed into helicity state in the Fourier space as
\begin{eqnarray}
\sigma_{i}(\eta, \bm{x}) &=& \int{\frac{{\rm d}^{3}k}{(2\pi)^3}}\sum_{\lambda = \pm 1}\sigma_{\lambda}(\eta, \bm{k})O^{(\lambda)}_{i}(\hat{\bm k})e^{-i\bm{k}\cdot \bm{x}} ~,
\end{eqnarray}
where $O^{(\lambda)}_{i}(\hat{\bm k})$ is the polarization vector defined in Refs.~\cite{Shiraishi:2010kd,Saga:2015apa}.
Note that because we work in standard cosmology based on the Boltzmann equation with Compton scattering and the Einstein equation, the parity symmetry does not break.
Therefore, the different helicity state of the vector mode takes the same evolution and variance, i.e., $\Braket{\sigma_{+}\sigma_{+}} = \Braket{\sigma_{-}\sigma_{-}}$.
The equation of motion for the second-order vector mode can be pulled from the second-order Einstein equation as
\begin{eqnarray}
\dot{\sigma}^{(2)}_{\lambda}(\eta ,{\bm k})+2\mathcal{H}\sigma^{(2)}_{\lambda}(\eta, {\bm k}) = {\cal S}^{(2)}_{\lambda}(\eta ,{\bm k}) ~, \label{eq: evolve sigma}
\end{eqnarray}
where a dot and $\mathcal{H}$ are the derivative with respect to the conformal time $\eta$ and the conformal Hubble parameter, respectively.
The right-hand side of the above Eq.~(\ref{eq: evolve sigma}) is the source term induced from the coupling of the first-order scalar mode.
The specific form of the source term ${\cal S}^{(2)}_{\lambda}(\eta ,{\bm k})$ is given in Ref.~\cite{Saga:2015apa}.
The dominant part of the source term is the quadratic of the scalar potential, naively, ${\cal S}^{(2)}_{\lambda} \sim \Phi^{(1)}\times \Phi^{(1)}$.
To compute the weak lensing signal, we need the dimensionless unequal-time power spectrum $\mathcal{P}_{V}(k, \eta, \eta')$ for the vector mode defined as
\begin{eqnarray}
\Braket{\sigma_{\lambda}(\eta, \bm{k})\sigma^{*}_{\lambda'}(\eta', \bm{k'})} &=& (2\pi)^{3}\delta_{\lambda,\lambda'}\delta^{3}_{\rm D}(\bm{k} - \bm{k'})\frac{2\pi^{2}}{k^{3}} \mathcal{P}_{V}(k, \eta, \eta') ~. \label{eq: Pk sigma}
\end{eqnarray}
The evolution equation for the second-order vector mode can be solved numerically.
Thus, we read the unequal-time power spectrum as the 21cm lensing curl mode.

Note that the power spectrum of the PGWs has the analytic solution during the matter-dominated eta as
\begin{eqnarray}
\mathcal{P}^{\rm PGW}_{T}(k, \eta, \eta') = 
r\Delta^2_{\mathcal{R}}(k_{0})\mathcal{T}_h^{({\rm PGW})}(k\eta )\mathcal{T}_h^{({\rm PGW})}(k\eta' ) ~, \label{eq: Pk h}
\end{eqnarray}
where $r$ is the tensor-to-scalar ratio which is constrained as $r \lesssim 0.1$.
Also $\Delta^2_{\mathcal{R}}(k_{0}) \approx 2.4\times 10^{-9}$ \cite{Hinshaw:2012aka} is the spectrum of the primordial curvature perturbation determined from cosmological observations.
The transfer function for PGWs $\mathcal{T}_h^{({\rm PGW})}(x)$ can be written by using the spherical Bessel function $j_{\ell}(x)$ as $\mathcal{T}_h^{({\rm PGW})}(x) = 3j_{1}(x)/x$.
Note that although this transfer function is the form during the matter-dominated era, the correction of the radiation to the spectrum of PGWs is small.
Therefore, we neglect its small contribution.
Throughout this paper, we fix the spectral index for the tensor mode to zero.

\subsection{Weak lensing curl mode}
In this subsection, we summarize the weak lensing curl mode and the lensing reconstruction \cite{Namikawa:2011cs}.
The curl mode is induced not from the scalar mode but from the vector and tensor modes.
The projected deflection angle on the celestial sphere $\Delta_{a}(\bm{\hat{n}})$ is decomposed into the scalar and pseudoscalar potentials by using the parity as \cite{Namikawa:2011cs}
\begin{equation}
\Delta_{a}(\bm{\hat{n}}) = \phi(\bm{\hat{n}})_{:a} + \varpi(\bm{\hat{n}})_{:b}\epsilon^{b}{}_{a}~, \label{eq: potentials}
\end{equation}
where we denote the scalar and pseudoscalar potentials as $\phi$ and $\varpi$, respectively.
The scalar and pseudoscalar potentials correspond to the weak lensing gradient and curl modes, respectively.
$\epsilon^{b}{}_{a}$ is the covariant two-dimensional Levi-Civit$\grave{\rm a}$ tensor and a colon is the covariant derivative on the unit sphere.
Note that latin characters started from $a, b, \cdots$ in the above equation denote $\theta$ and $\phi$.
From here, we drop the gradient mode since we are interested in the curl mode only.

By solving the perturbed geodesic equation in the Poisson gauge, we derive the solution of the projected deflection angle.
The solution can be written in terms of the metric perturbations as
\begin{eqnarray}
\varpi^{:a}{}_{:a} = - \int^{\chi_{\rm S}}_{0}d\chi \frac{\chi_{\rm S} - \chi}{\chi\chi_{\rm S}} \left[ \frac{d}{d\chi} \left( \chi \Omega^{a}{}_{:b} \epsilon^{b}{}_{a}\right)\right] ~,
\end{eqnarray}
where $\chi_{\rm S}$ is the comoving distance to the source.
In the case of CMB lensing, the comoving distance to the source corresponds to the last scattering surface.
On the other hand, in the case of 21cm lensing, there are many source redshifts depending on the observing frequency.
$\Omega_{a}$ consists of the vector and tensor metric perturbations as
\begin{equation}
\Omega_{a} = \left( -\sigma_{i} + h_{ij}e^{j}_{\chi}\right)e^{i}_{a}~,
\end{equation}
where $e^{i}_{\chi}$ and $e^{i}_{a}$ are the orthogonal spacelike basis along the light ray.

We expand the pseudoscalar potential by the spherical harmonics as
\begin{equation}
\varpi(\bm{\hat{n}}) = \sum_{\ell ,m} \varpi_{\ell ,m}Y_{\ell,m}(\bm{\hat{n}}) ~.
\end{equation}
Moreover, by using the above multipole coefficients, the angular power spectrum of the curl mode is defined as
\begin{equation}
C^{\varpi\varpi}_{\ell} = \frac{1}{2\ell + 1}\sum_{m=-\ell}^{\ell} \Braket{\varpi_{\ell ,m}\varpi^{*}_{\ell,m}} ~.
\end{equation}
The angular power spectra for the vector ($X=V$) and tensor ($X=T$) modes are related to the unequal-time power spectra of the vector and tensor metric perturbations as
\begin{eqnarray}
C^{(X)\varpi\varpi}_{\ell} = 
4\pi \int^{\infty}_{0} \frac{dk}{k}
\int^{\chi_{S}}_{0}{k d\chi}\int^{\chi_{S}}_{0}{k d\chi'}
\mathcal{S}^{(X)}_{\varpi, \ell}(k\chi)\mathcal{S}^{(X)}_{\varpi, \ell}(k\chi') \mathcal{P}_{X}(k, \eta_{0}-\chi,\eta_{0}-\chi') ~,
\label{eq: def cl}
\end{eqnarray}
where $\mathcal{S}^{(X)}_{\varpi, \ell}(k\chi)$ is the weight function defined as
\begin{eqnarray}
\mathcal{S}^{(V)}_{\varpi, \ell}(x) &=& \sqrt{\frac{(\ell-1)!}{(\ell+1)!}} \frac{j_{\ell}(x)}{x} ~, \\
\mathcal{S}^{(T)}_{\varpi, \ell}(x) &=& \frac{1}{2}\frac{(\ell-1)!}{(\ell+1)!}\sqrt{\frac{(\ell + 2)!}{(\ell -2)!}} \frac{j_{\ell}(x)}{x^{2}} ~.
\end{eqnarray}
The unequal-time power spectra for the vector and tensor modes are defined in Eqs.~(\ref{eq: Pk sigma}) and (\ref{eq: Pk h}).

Next, we review the 21cm lensing reconstruction for the curl mode following Ref.~\cite{Book:2011dz}.
Because of the difference in parity between the scalar and pseudoscalar potentials [see Eq.~(\ref{eq: potentials})], we can reconstruct the gradient and curl modes separately from the maps.
Throughout this paper, we assume that the detectability of the curl mode is based on the quadratic estimator as was used by the Planck collaboration.
The accuracy of the quadratic estimator is limited by the cosmic variance of the lensed CMB maps.
In our study, by the term ``the ideal experiment'' we mean that the reconstruction noise is due to the quadratic estimator without the instrumental noise.

First, as well as the CMB lensing reconstruction technique, we can reconstruct the curl mode from a single redshift slice.
In this case, the reconstruction noise is given as \cite{Book:2011dz,Namikawa:2011cs}
\begin{eqnarray}
N^{\varpi\varpi}_{\ell} &\equiv& \Braket{\left| n^{\varpi}_{\ell, m}\right|^{2}} \notag \\
&=& \left[ \frac{1}{2\ell + 1} \sum^{\ell_{\rm max}}_{L_{1},L_{2} = 2}f^{\varpi}_{\ell, L_{1},L_{2}}g^{\varpi}_{\ell, L_{1},L_{2}}\right]^{-1} ~, \label{eq: def noise}
\end{eqnarray}
where $f^{\varpi}_{\ell, L_{1},L_{2}}$ and $g^{\varpi}_{\ell, L_{1},L_{2}}$ can be expressed as follows:
\begin{eqnarray}
f^{\varpi}_{\ell, L_{1},L_{2}} &=& {}_{0}S^{\varpi}_{L_{1},\ell,L_{2}}C_{L_{2}} - {}_{0}S^{\varpi}_{L_{2},\ell,L_{1}}C_{L_{1}} ~, \\
{}_{0}S^{\varpi}_{L,\ell,\ell'} &=&
(-i)\sqrt{\frac{(2L+1)(2\ell+1)(2\ell' +1)}{16\pi}}
\sqrt{\ell(\ell+1)}\sqrt{\ell'(\ell'+1)}
\left[
\left(
\begin{array}{c c c}
L & \ell & \ell' \\
0 & -1 & 1 
\end{array}
\right)
- 
\left(
\begin{array}{c c c}
L & \ell & \ell' \\
0 & 1 & -1
\end{array}
\right)
 \right] ~, \\
 g^{\varpi}_{\ell, L_{1},L_{2}} &=&
\frac{\left( f^{\varpi}_{\ell,L_{1},L_{2}}\right)^{*}}{2\tilde{C}_{L}\tilde{C}_{L_{2}}} ~,
\end{eqnarray}
where $C_{\ell}$ and $\tilde{C}_{\ell}$ are the unlensed and lensed 21cm angular power spectra, respectively.
Note that due to the property of the Wigner-3j symbol, ${}_{0}S^{\varpi}_{L,\ell,\ell'} = 0$, when $L + \ell + \ell' = \mbox{even}$.
To discuss the detectability of the 21cm lensing curl mode, we introduce the signal-to-noise ratio as
\begin{eqnarray}
\left( \frac{{\rm S}}{\rm N}\right)^{\varpi\varpi}_{<\ell} = \left[ \sum_{\ell' = 2}^{\ell} \left( \frac{C^{\varpi\varpi}_{\ell'}}{\Delta C^{\varpi\varpi}_{\ell'}}\right)^{2}\right]^{1/2} ~, \label{eq: SN formula}
\end{eqnarray}
where we define the error as
\begin{eqnarray}
\Delta C^{\varpi\varpi}_{\ell} \equiv \sqrt{\frac{2}{2\ell + 1}}\left( C^{\varpi\varpi}_{\ell} + N^{\varpi\varpi}_{\ell}\right) ~.
\end{eqnarray}
Note that we assume an ideal experiment where the sky coverage fraction $f_{\rm sky}$ is unity.
The 21cm angular power spectrum can extend up to the multipole moments $\ell \approx 10^{6}\sim 10^{7}$ since there is no diffusion mechanism after the recombination era.
Therefore, even if we use a single redshift slice to reconstruct the 21cm curl mode, the noise spectrum from the 21cm angular power spectrum becomes smaller than that from the CMB angular power spectrum.

Moreover, we can further reduce the noise by coadding many redshift slices.
Following Ref.~\cite{Book:2011dz}, the number of the statistically independent redshift shells can be estimated as below.
The comoving distance between the neighboring statistically independent maps $\delta R$ is roughly related to the highest multipole moment $\ell_{\rm max}$ used in the lensing reconstruction as $\delta R \approx R \ell^{-1}_{\rm max}$, where $R$ is the comoving distance corresponding to the source redshift.
Therefore, the total number of available maps can be estimated as $\Delta R/\delta R \approx 0.15\ell_{\rm max}$, where $\Delta R$ is the comoving distance between $z_{\rm min}$ and $z_{\rm max}$.
Therefore if the lensing signal is mostly contributed from $z \lesssim 30$, the noise spectrum is drastically reduced by the factor of $0.15\ell_{\rm max}$.
In this paper, we call this reduced noise power spectrum the coadded noise spectrum.
In the following section, we present our main results and discussions.

\section{Results and Discussions}\label{sec: results}
\begin{figure}[t]
\begin{center}
\includegraphics[width=0.49\textwidth]{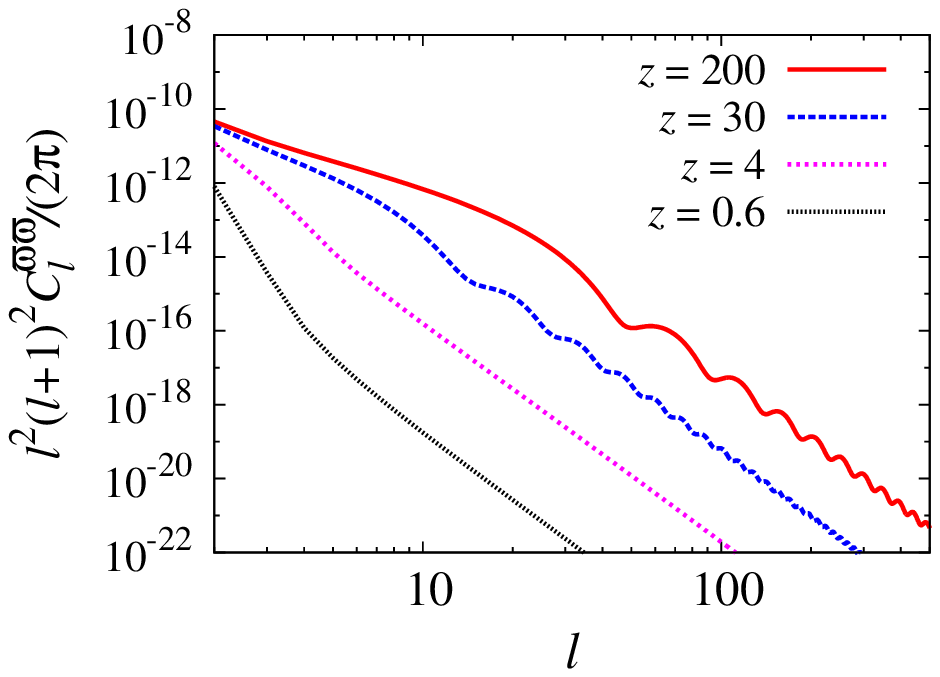}
\includegraphics[width=0.49\textwidth]{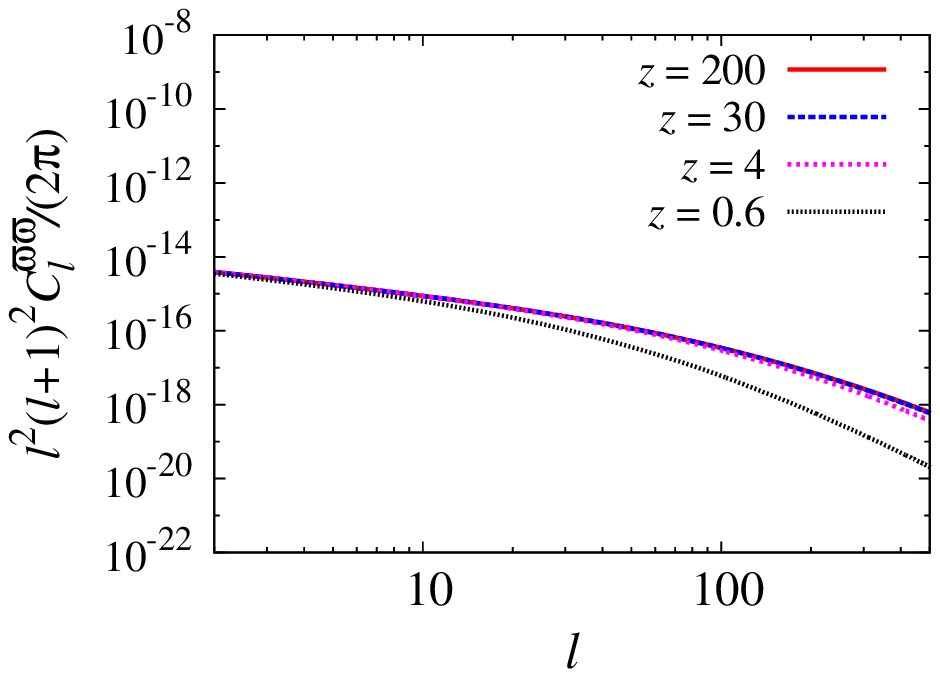}
\end{center}
\caption{%
The angular power spectra of the curl mode induced by PGWs with $r = 0.1$ (left) and the second-order vector mode (right) for redshifts from $z = 200$ to $0.6$ as indicated in the figures.
The curl mode from PGWs is substantially suppressed on small scales compared with that from the second-order vector mode.}
\label{fig: cls}
\end{figure}

In Fig.~\ref{fig: cls}, we show the angular power spectra of the curl mode induced from PGWs with $r = 0.1$ and the second-order vector mode.
We can see that the lensing signal from PGWs is suppressed as the redshift decreases.
On the other hand, the curl mode from the second-order vector mode remains almost constant.
We find that the redshift dependence of the second-order vector mode is similar to that of the gradient mode from the first-order scalar potential \cite{Lewis:2006fu}.
This is because the second-order vector mode is also sourced from the first-order scalar potential.
Therefore the amplitude of the second-order vector mode can have a greater amplitude than the curl mode from PGWs.
The amplitude of the curl mode from the second-order vector mode is greater than that from PGWs with $r = 0.1$ on smaller scales, such as $\ell \gtrsim 20$.
Furthermore, when the tensor-to-scalar ratio is quite small, e.g., $r \lesssim 10^{-5}$, the curl mode from the second-order vector mode dominates over almost all scales.
From this fact, we can conclude that even if we consider ideal observations, it would be difficult to hunt the tensor-to-scalar ratio $r \lesssim 10^{-5}$ by using 21cm lensing.

\begin{figure}[t]
\begin{center}
\includegraphics[width=0.49\textwidth]{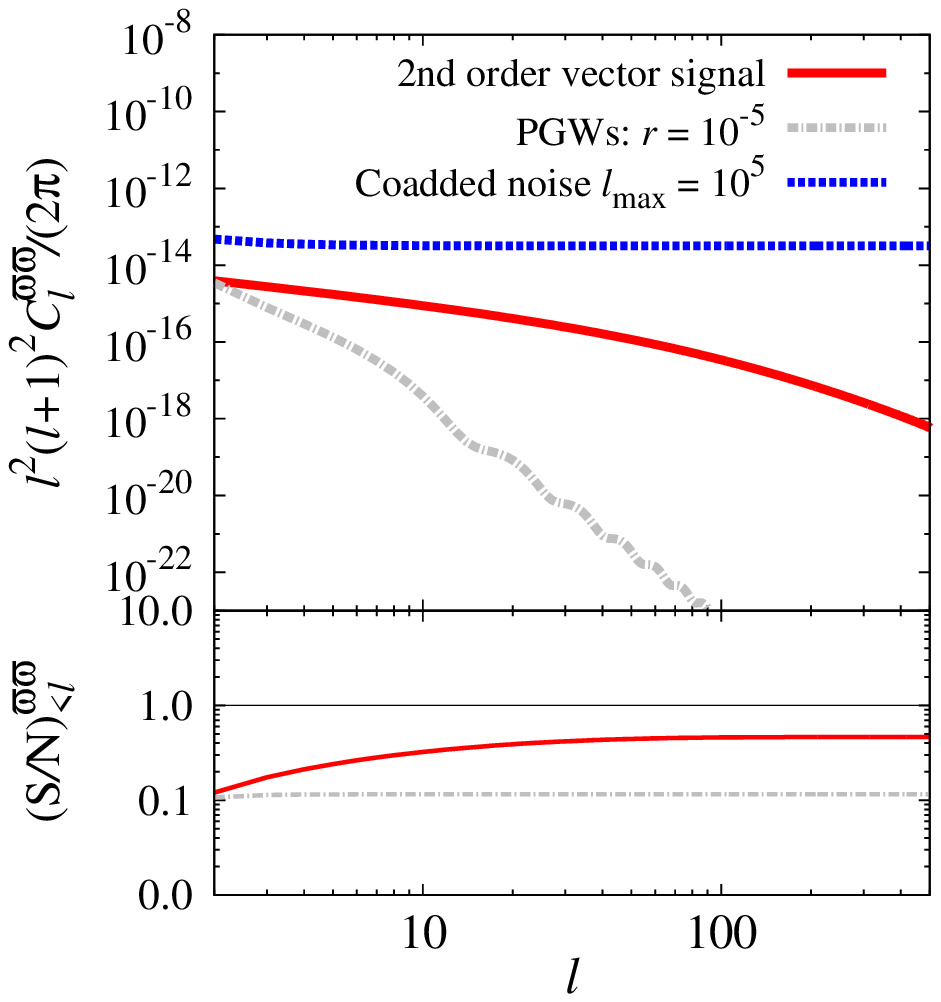}
\includegraphics[width=0.49\textwidth]{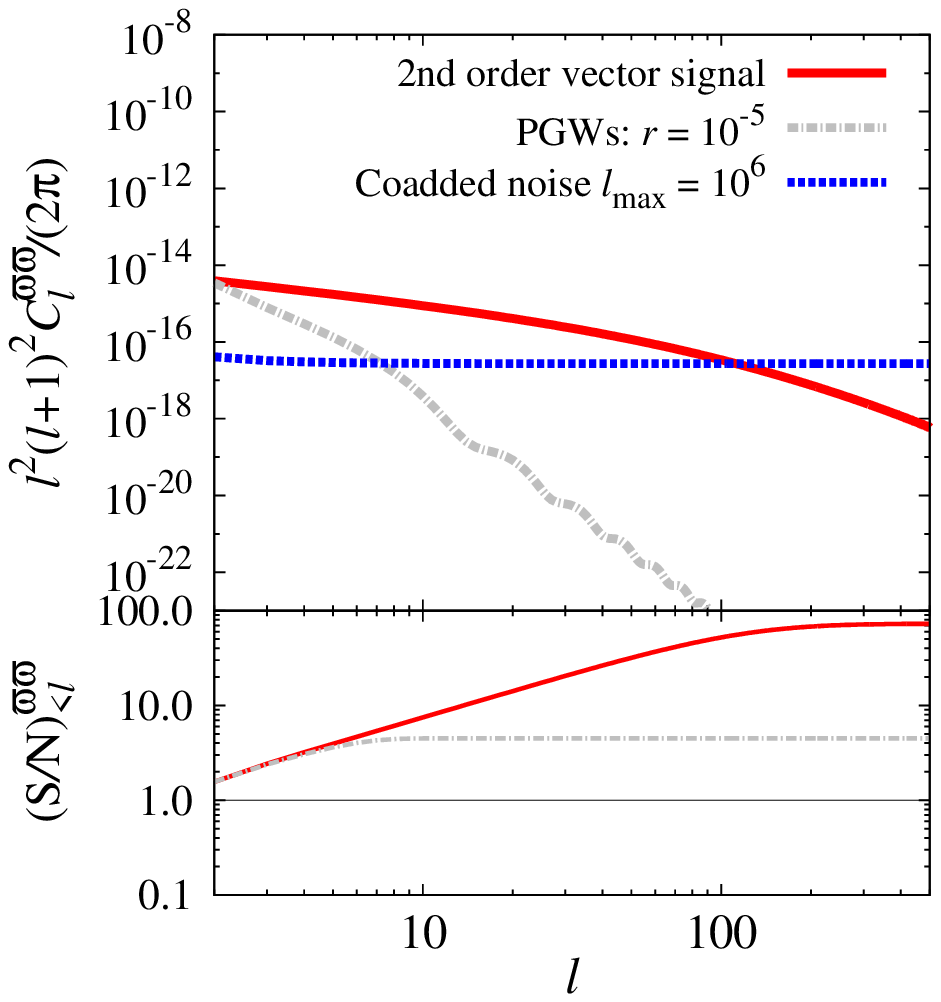}
\end{center}
\caption{%
The angular power spectrum of the curl mode from the second-order vector mode and the coadded reconstruction noise by using $\ell_{\rm max} = 10^{5}$ (top left) and $\ell_{\rm max} = 10^{6}$ (top right).
Bottom: The signal-to-noise ratio estimated by Eq.~(\ref{eq: SN formula}).
For reference, we also show the curl mode signal and signal-to-noise ratio induced by PGWs with $r = 10^{-5}$ and which tensor-to-scalar ratio corresponds to the same amplitude of the second-order curl mode at $\ell = 2$.
}
\label{fig: SN result}
\end{figure}

In Fig.~\ref{fig: SN result}, we depict the signal-to-noise ratio for two different values of $\ell_{\rm max} = 10^{5}$ and $10^{6}$, which is our main result in this paper.
For reference, we also show the signal-to-noise ratio for the case of PGWs with $r=10^{-5}$.
PGWs with $r = 10^{-5}$ have almost the same amplitude of the curl mode from the second-order vector mode at $\ell = 2$.
In the case of $\ell_{\rm max} = 10^{5}$, the signal-to-noise ratio reaches ${\rm S/N}\approx 0.11$ for the PGWs and ${\rm S/N}\approx 0.46$ for the second-order vector mode and it would be difficult to detect the second-order vector mode and PGWs with $r = 10^{-5}$.
On the other hand, in the case of $\ell_{\rm max} = 10^{6}$, we obtain ${\rm S/N}\approx 4.5$ for the PGWs and ${\rm S/N}\approx 73$ for the second-order vector mode.

The above signal-to-noise ratio is derived by adopting the reconstruction noise spectrum from the quadratic estimator performed in Refs.~\cite{Namikawa:2011cs,Ade:2015zua}.
Reconstruction noise from the quadratic estimator is limited by the cosmic variance of the lensed CMB fluctuations.
Ultimately, the iterative estimator proposed in Ref.~\cite{Hirata:2003ka} can reduce the reconstruction noise to zero.
Even in that case, the fact that the curl mode from PGW with $r\lesssim10^{-5}$ is concealed by that from the second-order vector mode does not change.

The signal-to-noise ratio of the second-order vector mode can reach higher than that of PGWs.
PGWs do not induce the curl mode amplitude on smaller scales since PGWs decay on subhorizon scales.
On the other hand, the second-order curl mode can remain large on smaller scales and at low redshift
since the second-order vector mode is continuously sourced by the first-order scalar gravitational potential.
The second-order vector mode grows on subhorizon scales.
From this nature, the second-order vector mode may be easier to detect than PGWs on small scales.

There is another source of the curl mode, that is, the lens-lens coupling examined in Refs.~\cite{Jain:1999ir,Sarkar:2008ii,Cooray:2002mj}.
The lens-lens coupling is sourced by the higher-order Born correction.
However, this correction mainly contributes the curl mode on small scales such as $\ell \gg 10$.
The curl mode on large scales is important to distinguish the PGWs and the second-order vector mode
since the PGWs and the second-order vector mode affect the curl mode on large scales, that is, $\ell \lesssim 10$.
When we consider the curl mode on all scales, the lens-lens coupling and the second-order vector mode should be taken into account.

To close this section, we describe a feature of the second-order vector mode.
The second-order vector mode does not have the free parameter since its source, that is, the first-order scalar mode, is well determined by the current cosmological observations.
Therefore, the prediction of the 21cm lensing curl mode from the second-order vector mode is quite robust.

\section{Summary}\label{sec: summary}
In this paper, we studied the detectability of the second-order vector mode by using 21cm radiation from the dark ages.
21cm radiation during the dark ages is a powerful tool to explore small signals such as second-order signatures since 21cm radiation anisotropy on small scales makes multipole moments available up to $\sim 10^{6}$.
Furthermore, by multifrequency observations, we can use many redshift slices to decrease the lensing reconstruction noise.
We focused on the weak lensing signal of the 21cm radiation from the dark ages.
As well as the CMB lensing, the 21cm photons are deflected by the foreground scalar, vector, and tensor perturbations.
The deflection angle of the 21cm photons can be decomposed into the scalar (gradient mode) and pseudoscalar (curl mode) potentials depending on its parity.
The curl mode is a good tracer of the cosmological vector and tensor modes since the scalar mode induces only the gradient mode.
It is known that the second-order tensor mode is the subdominant component in the large-scale structure signal such as weak lensing.
On the other hand, the second-order vector mode can have a comparable contribution on large-scale structure to primordial gravitational waves.
Accordingly, the observation that can detect PGWs with small tensor-to-scalar ratio can also be used to detect the second-order vector mode with high signal-to-noise ratio.

We discussed the detectability of the 21cm lensing curl mode induced from the second-order vector mode for the first time.
If the available multipole is limited to $\ell \lesssim 10^{5}$, the 21cm lensing curl mode from the second-order vector mode cannot be detected.
If we can extend the maximum multipole up to $\ell_{\rm max} \approx 10^{6}$, the signal-to-noise ratio reaches $73$.
We conclude that, in principle, we can explore the second-order vector mode by using 21cm radiation from the dark ages.
By comparing PGWs, it was also found that the PGWs with a tensor-to-scalar ratio $r \approx 10^{-5}$ become subdominant in the 21cm lensing curl mode.
In the previous study \cite{Book:2011dz}, they concluded that it is possible to detect the PGWs with $r\approx 10^{-9}$.
However, when second-order effects are included in their analysis, a tensor-to-scalar ratio smaller than $r\lesssim 10^{-5}$ would be difficult to detect by the 21cm lensing curl mode.
We can generalize this discussion for any model, including the vector or tensor modes with model parameters.
The second-order vector mode is generated from the first-order scalar mode that has been well determined by current observations.
Therefore, the 21cm curl mode from the second-order vector mode always exists.
Even if 21cm lensing is induced by other models, an amplitude smaller than the second-order vector mode is difficult to detect with 21cm lensing.

Throughout this paper, we assumed the ideal and challenging experiment for 21cm signals.
There are some forthcoming observations for 21cm signals after recombination, e.g., from the Square Kilometer Array.
Moreover, exploring 21cm radiation must become an active topic in the near future.
Before starting these observations, exploring the potentials of 21cm radiation is important and this work gives one of the nontrivial solutions.

\begin{acknowledgments}
I thank Kiyotomo Ichiki and Hiroyuki Tashiro for useful comments and discussions.
This work was supported in part by a Grant-in-Aid for JSPS Research under Grant No.~26-63 (S.S.).
I also acknowledge the Kobayashi-Maskawa Institute for the Origin of Particles and the Universe, Nagoya University, for providing useful computing resources for conducting this research.
\end{acknowledgments}
\bibliography{ref}

\begin{thebibliography}{37}
\expandafter\ifx\csname natexlab\endcsname\relax\def\natexlab#1{#1}\fi
\expandafter\ifx\csname bibnamefont\endcsname\relax
  \def\bibnamefont#1{#1}\fi
\expandafter\ifx\csname bibfnamefont\endcsname\relax
  \def\bibfnamefont#1{#1}\fi
\expandafter\ifx\csname citenamefont\endcsname\relax
  \def\citenamefont#1{#1}\fi
\expandafter\ifx\csname url\endcsname\relax
  \def\url#1{\texttt{#1}}\fi
\expandafter\ifx\csname urlprefix\endcsname\relax\def\urlprefix{URL }\fi
\providecommand{\bibinfo}[2]{#2}
\providecommand{\eprint}[2][]{\url{#2}}

\bibitem[{\citenamefont{Tegmark et~al.}(2006)}]{Tegmark:2006az}
\bibinfo{author}{\bibfnamefont{M.}~\bibnamefont{Tegmark}} \bibnamefont{et~al.}
  (\bibinfo{collaboration}{SDSS}), \bibinfo{journal}{Phys. Rev.}
  \textbf{\bibinfo{volume}{D74}}, \bibinfo{pages}{123507}
  (\bibinfo{year}{2006}), \eprint{astro-ph/0608632}.

\bibitem[{\citenamefont{Hinshaw et~al.}(2013)}]{Hinshaw:2012aka}
\bibinfo{author}{\bibfnamefont{G.}~\bibnamefont{Hinshaw}} \bibnamefont{et~al.}
  (\bibinfo{collaboration}{WMAP}), \bibinfo{journal}{Astrophys. J. Suppl.}
  \textbf{\bibinfo{volume}{208}}, \bibinfo{pages}{19} (\bibinfo{year}{2013}),
  \eprint{1212.5226}.

\bibitem[{\citenamefont{Sanchez et~al.}(2014)}]{Sanchez:2013tga}
\bibinfo{author}{\bibfnamefont{A.~G.} \bibnamefont{Sanchez}}
  \bibnamefont{et~al.}, \bibinfo{journal}{Mon. Not. Roy. Astron. Soc.}
  \textbf{\bibinfo{volume}{440}}, \bibinfo{pages}{2692} (\bibinfo{year}{2014}),
  \eprint{1312.4854}.

\bibitem[{\citenamefont{Ade et~al.}(2015{\natexlab{a}})}]{Ade:2015xua}
\bibinfo{author}{\bibfnamefont{P.~A.~R.} \bibnamefont{Ade}}
  \bibnamefont{et~al.} (\bibinfo{collaboration}{Planck})
  (\bibinfo{year}{2015}{\natexlab{a}}), \eprint{1502.01589}.

\bibitem[{\citenamefont{Lewis}(2004{\natexlab{a}})}]{Lewis:2004ef}
\bibinfo{author}{\bibfnamefont{A.}~\bibnamefont{Lewis}},
  \bibinfo{journal}{Phys. Rev.} \textbf{\bibinfo{volume}{D70}},
  \bibinfo{pages}{043011} (\bibinfo{year}{2004}{\natexlab{a}}),
  \eprint{astro-ph/0406096}.

\bibitem[{\citenamefont{Lewis}(2004{\natexlab{b}})}]{Lewis:2004kg}
\bibinfo{author}{\bibfnamefont{A.}~\bibnamefont{Lewis}},
  \bibinfo{journal}{Phys. Rev.} \textbf{\bibinfo{volume}{D70}},
  \bibinfo{pages}{043518} (\bibinfo{year}{2004}{\natexlab{b}}),
  \eprint{astro-ph/0403583}.

\bibitem[{\citenamefont{Pen et~al.}(1997)\citenamefont{Pen, Seljak, and
  Turok}}]{Pen:1997ae}
\bibinfo{author}{\bibfnamefont{U.-L.} \bibnamefont{Pen}},
  \bibinfo{author}{\bibfnamefont{U.}~\bibnamefont{Seljak}}, \bibnamefont{and}
  \bibinfo{author}{\bibfnamefont{N.}~\bibnamefont{Turok}},
  \bibinfo{journal}{Phys. Rev. Lett.} \textbf{\bibinfo{volume}{79}},
  \bibinfo{pages}{1611} (\bibinfo{year}{1997}), \eprint{astro-ph/9704165}.

\bibitem[{\citenamefont{Durrer et~al.}(1999)\citenamefont{Durrer, Kunz, and
  Melchiorri}}]{Durrer:1998rw}
\bibinfo{author}{\bibfnamefont{R.}~\bibnamefont{Durrer}},
  \bibinfo{author}{\bibfnamefont{M.}~\bibnamefont{Kunz}}, \bibnamefont{and}
  \bibinfo{author}{\bibfnamefont{A.}~\bibnamefont{Melchiorri}},
  \bibinfo{journal}{Phys. Rev.} \textbf{\bibinfo{volume}{D59}},
  \bibinfo{pages}{123005} (\bibinfo{year}{1999}), \eprint{astro-ph/9811174}.

\bibitem[{\citenamefont{Horiguchi et~al.}(2015)\citenamefont{Horiguchi, Ichiki,
  Sekiguchi, and Sugiyama}}]{Horiguchi:2015xsa}
\bibinfo{author}{\bibfnamefont{K.}~\bibnamefont{Horiguchi}},
  \bibinfo{author}{\bibfnamefont{K.}~\bibnamefont{Ichiki}},
  \bibinfo{author}{\bibfnamefont{T.}~\bibnamefont{Sekiguchi}},
  \bibnamefont{and} \bibinfo{author}{\bibfnamefont{N.}~\bibnamefont{Sugiyama}},
  \bibinfo{journal}{JCAP} \textbf{\bibinfo{volume}{1504}}, \bibinfo{pages}{007}
  (\bibinfo{year}{2015}), \eprint{1501.06304}.

\bibitem[{\citenamefont{Zuntz et~al.}(2010)\citenamefont{Zuntz, Zlosnik,
  Bourliot, Ferreira, and Starkman}}]{Zuntz:2010jp}
\bibinfo{author}{\bibfnamefont{J.}~\bibnamefont{Zuntz}},
  \bibinfo{author}{\bibfnamefont{T.~G.} \bibnamefont{Zlosnik}},
  \bibinfo{author}{\bibfnamefont{F.}~\bibnamefont{Bourliot}},
  \bibinfo{author}{\bibfnamefont{P.~G.} \bibnamefont{Ferreira}},
  \bibnamefont{and} \bibinfo{author}{\bibfnamefont{G.~D.}
  \bibnamefont{Starkman}}, \bibinfo{journal}{Phys. Rev.}
  \textbf{\bibinfo{volume}{D81}}, \bibinfo{pages}{104015}
  (\bibinfo{year}{2010}), \eprint{1002.0849}.

\bibitem[{\citenamefont{Saga et~al.}(2013)\citenamefont{Saga, Shiraishi,
  Ichiki, and Sugiyama}}]{Saga:2013glg}
\bibinfo{author}{\bibfnamefont{S.}~\bibnamefont{Saga}},
  \bibinfo{author}{\bibfnamefont{M.}~\bibnamefont{Shiraishi}},
  \bibinfo{author}{\bibfnamefont{K.}~\bibnamefont{Ichiki}}, \bibnamefont{and}
  \bibinfo{author}{\bibfnamefont{N.}~\bibnamefont{Sugiyama}},
  \bibinfo{journal}{Phys. Rev.} \textbf{\bibinfo{volume}{D87}},
  \bibinfo{pages}{104025} (\bibinfo{year}{2013}), \eprint{1302.4189}.

\bibitem[{\citenamefont{Assadullahi and Wands}(2010)}]{Assadullahi:2009jc}
\bibinfo{author}{\bibfnamefont{H.}~\bibnamefont{Assadullahi}} \bibnamefont{and}
  \bibinfo{author}{\bibfnamefont{D.}~\bibnamefont{Wands}},
  \bibinfo{journal}{Phys. Rev.} \textbf{\bibinfo{volume}{D81}},
  \bibinfo{pages}{023527} (\bibinfo{year}{2010}), \eprint{0907.4073}.

\bibitem[{\citenamefont{Ananda et~al.}(2007)\citenamefont{Ananda, Clarkson, and
  Wands}}]{Ananda:2006af}
\bibinfo{author}{\bibfnamefont{K.~N.} \bibnamefont{Ananda}},
  \bibinfo{author}{\bibfnamefont{C.}~\bibnamefont{Clarkson}}, \bibnamefont{and}
  \bibinfo{author}{\bibfnamefont{D.}~\bibnamefont{Wands}},
  \bibinfo{journal}{Phys. Rev.} \textbf{\bibinfo{volume}{D75}},
  \bibinfo{pages}{123518} (\bibinfo{year}{2007}), \eprint{gr-qc/0612013}.

\bibitem[{\citenamefont{Saga et~al.}(2015{\natexlab{a}})\citenamefont{Saga,
  Ichiki, and Sugiyama}}]{Saga:2014jca}
\bibinfo{author}{\bibfnamefont{S.}~\bibnamefont{Saga}},
  \bibinfo{author}{\bibfnamefont{K.}~\bibnamefont{Ichiki}}, \bibnamefont{and}
  \bibinfo{author}{\bibfnamefont{N.}~\bibnamefont{Sugiyama}},
  \bibinfo{journal}{Phys. Rev.} \textbf{\bibinfo{volume}{D91}},
  \bibinfo{pages}{024030} (\bibinfo{year}{2015}{\natexlab{a}}),
  \eprint{1412.1081}.

\bibitem[{\citenamefont{Baumann et~al.}(2007)\citenamefont{Baumann, Steinhardt,
  Takahashi, and Ichiki}}]{Baumann:2007zm}
\bibinfo{author}{\bibfnamefont{D.}~\bibnamefont{Baumann}},
  \bibinfo{author}{\bibfnamefont{P.~J.} \bibnamefont{Steinhardt}},
  \bibinfo{author}{\bibfnamefont{K.}~\bibnamefont{Takahashi}},
  \bibnamefont{and} \bibinfo{author}{\bibfnamefont{K.}~\bibnamefont{Ichiki}},
  \bibinfo{journal}{Phys. Rev.} \textbf{\bibinfo{volume}{D76}},
  \bibinfo{pages}{084019} (\bibinfo{year}{2007}), \eprint{hep-th/0703290}.

\bibitem[{\citenamefont{Saga et~al.}(2015{\natexlab{b}})\citenamefont{Saga,
  Yamauchi, and Ichiki}}]{Saga:2015apa}
\bibinfo{author}{\bibfnamefont{S.}~\bibnamefont{Saga}},
  \bibinfo{author}{\bibfnamefont{D.}~\bibnamefont{Yamauchi}}, \bibnamefont{and}
  \bibinfo{author}{\bibfnamefont{K.}~\bibnamefont{Ichiki}},
  \bibinfo{journal}{Phys. Rev.} \textbf{\bibinfo{volume}{D92}},
  \bibinfo{pages}{063533} (\bibinfo{year}{2015}{\natexlab{b}}),
  \eprint{1505.02774}.

\bibitem[{\citenamefont{Ichiki et~al.}(2006)\citenamefont{Ichiki, Takahashi,
  Ohno, Hanayama, and Sugiyama}}]{Ichiki:2006cd}
\bibinfo{author}{\bibfnamefont{K.}~\bibnamefont{Ichiki}},
  \bibinfo{author}{\bibfnamefont{K.}~\bibnamefont{Takahashi}},
  \bibinfo{author}{\bibfnamefont{H.}~\bibnamefont{Ohno}},
  \bibinfo{author}{\bibfnamefont{H.}~\bibnamefont{Hanayama}}, \bibnamefont{and}
  \bibinfo{author}{\bibfnamefont{N.}~\bibnamefont{Sugiyama}},
  \bibinfo{journal}{Science} \textbf{\bibinfo{volume}{311}},
  \bibinfo{pages}{827} (\bibinfo{year}{2006}), \eprint{astro-ph/0603631}.

\bibitem[{\citenamefont{Fenu et~al.}(2011)\citenamefont{Fenu, Pitrou, and
  Maartens}}]{Fenu:2010kh}
\bibinfo{author}{\bibfnamefont{E.}~\bibnamefont{Fenu}},
  \bibinfo{author}{\bibfnamefont{C.}~\bibnamefont{Pitrou}}, \bibnamefont{and}
  \bibinfo{author}{\bibfnamefont{R.}~\bibnamefont{Maartens}},
  \bibinfo{journal}{Mon. Not. Roy. Astron. Soc.}
  \textbf{\bibinfo{volume}{414}}, \bibinfo{pages}{2354} (\bibinfo{year}{2011}),
  \eprint{1012.2958}.

\bibitem[{\citenamefont{Saga et~al.}(2015{\natexlab{c}})\citenamefont{Saga,
  Ichiki, Takahashi, and Sugiyama}}]{Saga:2015bna}
\bibinfo{author}{\bibfnamefont{S.}~\bibnamefont{Saga}},
  \bibinfo{author}{\bibfnamefont{K.}~\bibnamefont{Ichiki}},
  \bibinfo{author}{\bibfnamefont{K.}~\bibnamefont{Takahashi}},
  \bibnamefont{and} \bibinfo{author}{\bibfnamefont{N.}~\bibnamefont{Sugiyama}},
  \bibinfo{journal}{Phys. Rev.} \textbf{\bibinfo{volume}{D91}},
  \bibinfo{pages}{123510} (\bibinfo{year}{2015}{\natexlab{c}}),
  \eprint{1504.03790}.

\bibitem[{\citenamefont{Namikawa et~al.}(2012)\citenamefont{Namikawa, Yamauchi,
  and Taruya}}]{Namikawa:2011cs}
\bibinfo{author}{\bibfnamefont{T.}~\bibnamefont{Namikawa}},
  \bibinfo{author}{\bibfnamefont{D.}~\bibnamefont{Yamauchi}}, \bibnamefont{and}
  \bibinfo{author}{\bibfnamefont{A.}~\bibnamefont{Taruya}},
  \bibinfo{journal}{JCAP} \textbf{\bibinfo{volume}{1201}}, \bibinfo{pages}{007}
  (\bibinfo{year}{2012}), \eprint{1110.1718}.

\bibitem[{\citenamefont{Yamauchi et~al.}(2012)\citenamefont{Yamauchi, Namikawa,
  and Taruya}}]{Yamauchi:2012bc}
\bibinfo{author}{\bibfnamefont{D.}~\bibnamefont{Yamauchi}},
  \bibinfo{author}{\bibfnamefont{T.}~\bibnamefont{Namikawa}}, \bibnamefont{and}
  \bibinfo{author}{\bibfnamefont{A.}~\bibnamefont{Taruya}},
  \bibinfo{journal}{JCAP} \textbf{\bibinfo{volume}{1210}}, \bibinfo{pages}{030}
  (\bibinfo{year}{2012}), \eprint{1205.2139}.

\bibitem[{\citenamefont{Yamauchi et~al.}(2013)\citenamefont{Yamauchi, Namikawa,
  and Taruya}}]{Yamauchi:2013fra}
\bibinfo{author}{\bibfnamefont{D.}~\bibnamefont{Yamauchi}},
  \bibinfo{author}{\bibfnamefont{T.}~\bibnamefont{Namikawa}}, \bibnamefont{and}
  \bibinfo{author}{\bibfnamefont{A.}~\bibnamefont{Taruya}},
  \bibinfo{journal}{JCAP} \textbf{\bibinfo{volume}{1308}}, \bibinfo{pages}{051}
  (\bibinfo{year}{2013}), \eprint{1305.3348}.

\bibitem[{\citenamefont{Book et~al.}(2012)\citenamefont{Book, Kamionkowski, and
  Schmidt}}]{Book:2011dz}
\bibinfo{author}{\bibfnamefont{L.}~\bibnamefont{Book}},
  \bibinfo{author}{\bibfnamefont{M.}~\bibnamefont{Kamionkowski}},
  \bibnamefont{and} \bibinfo{author}{\bibfnamefont{F.}~\bibnamefont{Schmidt}},
  \bibinfo{journal}{Phys. Rev. Lett.} \textbf{\bibinfo{volume}{108}},
  \bibinfo{pages}{211301} (\bibinfo{year}{2012}), \eprint{1112.0567}.

\bibitem[{\citenamefont{Masui and Pen}(2010)}]{Masui:2010cz}
\bibinfo{author}{\bibfnamefont{K.~W.} \bibnamefont{Masui}} \bibnamefont{and}
  \bibinfo{author}{\bibfnamefont{U.-L.} \bibnamefont{Pen}},
  \bibinfo{journal}{Phys. Rev. Lett.} \textbf{\bibinfo{volume}{105}},
  \bibinfo{pages}{161302} (\bibinfo{year}{2010}), \eprint{1006.4181}.

\bibitem[{\citenamefont{Sigurdson and Cooray}(2005)}]{Sigurdson:2005cp}
\bibinfo{author}{\bibfnamefont{K.}~\bibnamefont{Sigurdson}} \bibnamefont{and}
  \bibinfo{author}{\bibfnamefont{A.}~\bibnamefont{Cooray}},
  \bibinfo{journal}{Phys. Rev. Lett.} \textbf{\bibinfo{volume}{95}},
  \bibinfo{pages}{211303} (\bibinfo{year}{2005}), \eprint{astro-ph/0502549}.

\bibitem[{\citenamefont{Furlanetto et~al.}(2006)\citenamefont{Furlanetto, Oh,
  and Briggs}}]{Furlanetto:2006jb}
\bibinfo{author}{\bibfnamefont{S.}~\bibnamefont{Furlanetto}},
  \bibinfo{author}{\bibfnamefont{S.~P.} \bibnamefont{Oh}}, \bibnamefont{and}
  \bibinfo{author}{\bibfnamefont{F.}~\bibnamefont{Briggs}},
  \bibinfo{journal}{Phys. Rept.} \textbf{\bibinfo{volume}{433}},
  \bibinfo{pages}{181} (\bibinfo{year}{2006}), \eprint{astro-ph/0608032}.

\bibitem[{\citenamefont{Lewis and Challinor}(2007)}]{Lewis:2007kz}
\bibinfo{author}{\bibfnamefont{A.}~\bibnamefont{Lewis}} \bibnamefont{and}
  \bibinfo{author}{\bibfnamefont{A.}~\bibnamefont{Challinor}},
  \bibinfo{journal}{Phys. Rev.} \textbf{\bibinfo{volume}{D76}},
  \bibinfo{pages}{083005} (\bibinfo{year}{2007}), \eprint{astro-ph/0702600}.

\bibitem[{\citenamefont{Bruni et~al.}(2014)\citenamefont{Bruni, Thomas, and
  Wands}}]{Bruni:2013mua}
\bibinfo{author}{\bibfnamefont{M.}~\bibnamefont{Bruni}},
  \bibinfo{author}{\bibfnamefont{D.~B.} \bibnamefont{Thomas}},
  \bibnamefont{and} \bibinfo{author}{\bibfnamefont{D.}~\bibnamefont{Wands}},
  \bibinfo{journal}{Phys. Rev.} \textbf{\bibinfo{volume}{D89}},
  \bibinfo{pages}{044010} (\bibinfo{year}{2014}), \eprint{1306.1562}.

\bibitem[{\citenamefont{Thomas et~al.}(2015)\citenamefont{Thomas, Bruni, and
  Wands}}]{Thomas:2014aga}
\bibinfo{author}{\bibfnamefont{D.~B.} \bibnamefont{Thomas}},
  \bibinfo{author}{\bibfnamefont{M.}~\bibnamefont{Bruni}}, \bibnamefont{and}
  \bibinfo{author}{\bibfnamefont{D.}~\bibnamefont{Wands}},
  \bibinfo{journal}{JCAP} \textbf{\bibinfo{volume}{1509}}, \bibinfo{pages}{021}
  (\bibinfo{year}{2015}), \eprint{1403.4947}.

\bibitem[{\citenamefont{Adamek et~al.}(2016)\citenamefont{Adamek, Durrer, and
  Tansella}}]{Adamek:2015mna}
\bibinfo{author}{\bibfnamefont{J.}~\bibnamefont{Adamek}},
  \bibinfo{author}{\bibfnamefont{R.}~\bibnamefont{Durrer}}, \bibnamefont{and}
  \bibinfo{author}{\bibfnamefont{V.}~\bibnamefont{Tansella}},
  \bibinfo{journal}{JCAP} \textbf{\bibinfo{volume}{1601}}, \bibinfo{pages}{024}
  (\bibinfo{year}{2016}), \eprint{1510.01566}.

\bibitem[{\citenamefont{Shiraishi et~al.}(2011)\citenamefont{Shiraishi, Nitta,
  Yokoyama, Ichiki, and Takahashi}}]{Shiraishi:2010kd}
\bibinfo{author}{\bibfnamefont{M.}~\bibnamefont{Shiraishi}},
  \bibinfo{author}{\bibfnamefont{D.}~\bibnamefont{Nitta}},
  \bibinfo{author}{\bibfnamefont{S.}~\bibnamefont{Yokoyama}},
  \bibinfo{author}{\bibfnamefont{K.}~\bibnamefont{Ichiki}}, \bibnamefont{and}
  \bibinfo{author}{\bibfnamefont{K.}~\bibnamefont{Takahashi}},
  \bibinfo{journal}{Prog. Theor. Phys.} \textbf{\bibinfo{volume}{125}},
  \bibinfo{pages}{795} (\bibinfo{year}{2011}), \eprint{1012.1079}.

\bibitem[{\citenamefont{Lewis and Challinor}(2006)}]{Lewis:2006fu}
\bibinfo{author}{\bibfnamefont{A.}~\bibnamefont{Lewis}} \bibnamefont{and}
  \bibinfo{author}{\bibfnamefont{A.}~\bibnamefont{Challinor}},
  \bibinfo{journal}{Phys. Rept.} \textbf{\bibinfo{volume}{429}},
  \bibinfo{pages}{1} (\bibinfo{year}{2006}), \eprint{astro-ph/0601594}.

\bibitem[{\citenamefont{Ade et~al.}(2015{\natexlab{b}})}]{Ade:2015zua}
\bibinfo{author}{\bibfnamefont{P.~A.~R.} \bibnamefont{Ade}}
  \bibnamefont{et~al.} (\bibinfo{collaboration}{Planck})
  (\bibinfo{year}{2015}{\natexlab{b}}), \eprint{1502.01591}.

\bibitem[{\citenamefont{Hirata and Seljak}(2003)}]{Hirata:2003ka}
\bibinfo{author}{\bibfnamefont{C.~M.} \bibnamefont{Hirata}} \bibnamefont{and}
  \bibinfo{author}{\bibfnamefont{U.}~\bibnamefont{Seljak}},
  \bibinfo{journal}{Phys. Rev.} \textbf{\bibinfo{volume}{D68}},
  \bibinfo{pages}{083002} (\bibinfo{year}{2003}), \eprint{astro-ph/0306354}.

\bibitem[{\citenamefont{Jain et~al.}(2000)\citenamefont{Jain, Seljak, and
  White}}]{Jain:1999ir}
\bibinfo{author}{\bibfnamefont{B.}~\bibnamefont{Jain}},
  \bibinfo{author}{\bibfnamefont{U.}~\bibnamefont{Seljak}}, \bibnamefont{and}
  \bibinfo{author}{\bibfnamefont{S.~D.~M.} \bibnamefont{White}},
  \bibinfo{journal}{Astrophys. J.} \textbf{\bibinfo{volume}{530}},
  \bibinfo{pages}{547} (\bibinfo{year}{2000}), \eprint{astro-ph/9901191}.

\bibitem[{\citenamefont{Sarkar et~al.}(2008)\citenamefont{Sarkar, Serra,
  Cooray, Ichiki, and Baumann}}]{Sarkar:2008ii}
\bibinfo{author}{\bibfnamefont{D.}~\bibnamefont{Sarkar}},
  \bibinfo{author}{\bibfnamefont{P.}~\bibnamefont{Serra}},
  \bibinfo{author}{\bibfnamefont{A.}~\bibnamefont{Cooray}},
  \bibinfo{author}{\bibfnamefont{K.}~\bibnamefont{Ichiki}}, \bibnamefont{and}
  \bibinfo{author}{\bibfnamefont{D.}~\bibnamefont{Baumann}},
  \bibinfo{journal}{Phys. Rev.} \textbf{\bibinfo{volume}{D77}},
  \bibinfo{pages}{103515} (\bibinfo{year}{2008}), \eprint{0803.1490}.

\bibitem[{\citenamefont{Cooray and Hu}(2002)}]{Cooray:2002mj}
\bibinfo{author}{\bibfnamefont{A.}~\bibnamefont{Cooray}} \bibnamefont{and}
  \bibinfo{author}{\bibfnamefont{W.}~\bibnamefont{Hu}},
  \bibinfo{journal}{Astrophys. J.} \textbf{\bibinfo{volume}{574}},
  \bibinfo{pages}{19} (\bibinfo{year}{2002}), \eprint{astro-ph/0202411}.

\end{thebibliography}
\end{document}